\documentclass[twocolumn,reprint,showpacs,aps,prl]{revtex4-1}
\usepackage{threeparttable}
\usepackage{amsmath}
\usepackage{lipsum}
\usepackage{multirow}
\usepackage{booktabs}
\usepackage{graphicx,color}
\usepackage{graphicx}
\usepackage{subfigure}
\usepackage{amsmath}

\newcommand{\ba}{\begin{eqnarray}}
\newcommand{\ea}{\end{eqnarray}}

\newcommand{\be}{\begin{equation}}
\newcommand{\ee}{\end{equation}}

\newcommand{\et}{{\it et al. }}

\definecolor{pink}{rgb}{1,0.18,1.0}
\def\prl{{ Phys. Rev. Lett. }}

\def\jcp{{ J. Chem. Phys. }}

\def\sci{{ Science }}

\def\jap{{J. Appl. Phys. }}

\def\nm{{Nature Mater. }}
\def\jpcm{{J. Phys.: Condens. Matter }}
\def\nrl{{Nanoscale Res. Lett. }}

\def\rmp{{Rev. Mod. Phys. }}

\def\nrl{{Nanoscale Res. Lett. }}
\def\prx{{Phys. Rev. X }}
\def\nl{{Nano Lett. }}

\begin{document}

\title{A theoretical study of blue phosphorene nanoribbons based on
  first-principles calculations}

\author{Jiafeng Xie}
\affiliation{Key Laboratory for Magnetism and Magnetic Materials of
 the Ministry of Education, Lanzhou University, Lanzhou 730000, China}

\author{M. S. Si$^{*}$}
\affiliation{Key Laboratory for Magnetism and Magnetic Materials of
 the Ministry of Education, Lanzhou University, Lanzhou 730000, China}

\author{D. Z. Yang}
\affiliation{Key Laboratory for Magnetism and Magnetic Materials of
 the Ministry of Education, Lanzhou University, Lanzhou 730000, China}

\author{Z. Y. Zhang}
\affiliation{Key Laboratory for Magnetism and Magnetic Materials of
 the Ministry of Education, Lanzhou University, Lanzhou 730000, China}

\author{D. S. Xue}
\affiliation{Key Laboratory for Magnetism and Magnetic Materials of
 the Ministry of Education, Lanzhou University, Lanzhou 730000, China}

\date{\today}

\begin{abstract}
Based on first-principles calculations, we present a quantum
  confinement mechanism for the band gaps of blue phosphorene
  nanoribbons (BPNRs) as a function of their widths. The BPNRs
  considered have either armchair or zigzag shaped edges on both sides
  with hydrogen saturation. Both the two types of nanoribbons are shown to be
  indirect semiconductors. An enhanced energy gap of around 1 eV can
  be realized when   the width decreases to $\sim$10 {\AA}. The underlying physics is ascribed
  to the quantum confinement. More importantly, the quantum
  confinement parameters are obtained by fitting the calculated
  gaps with respect to their widths. The results show that the quantum
  confinement in armchair nanoribbons is stronger than that in zigzag
  ones. This study provides an efficient approach to tune the 
  energy gap in BPNRs.  
\end{abstract}



\maketitle

\section{I. introduction}

Since the successful synthesis \cite{novoselov,zhang,berger} of
monolayer graphite, i.e., graphene, 
significant efforts have been invested to understand its
physical, chemical, thermal and other properties
\cite{neto,stankovich,nair,ferrari,zhang,geim,novoselov1}. 
One most enticing among them
is the high mobility of carrier confined within the two-dimensional (2D)
surface \cite{reich}. This makes graphene a good candidate in
semiconductor industry. However, it lacks a natural band gap, limiting its practical
applications. The electronic band gap is an intrinsic property of
semiconductors that determines their transport and governs the
operation of semiconductor devices. 
To this end, many other 2D materials are explored 
theoretically and experimentally
\cite{corso,simsjcp,niunrl,zhangjcp,silicene,prx}. 
A search for an intrinsic energy gap is
prerequisite to allow the efficient control of carriers by external
field.

Unlike graphene, black phosphorus in its bulk phase is an intrinsic
semiconductor with a energy gap of around 0.19 eV \cite{du}. In
theory, it can be isolated to single layer as the weak van der Waals
interaction links the layers together. With this respect, two groups
\cite{liu,li} independently reported that they stripped black phosphorus
to few layers. Li \et further predicted that the energy gap can be
increased to be $\sim$0.8 eV as black phosphorus decreases to one layer
\cite{li}. At the same time, a strain of 10\% can also introduce a
considerable energy gap of about 1 eV \cite{li,rodin}. But, the
uncontrollable strain is a challenge in
experiment.  Unfortunately, these gap values are still much smaller
than the ideal value of $\sim$2 eV, which is commonly used in
electronic devices.

\begin{figure}
\includegraphics[width=7.5cm]{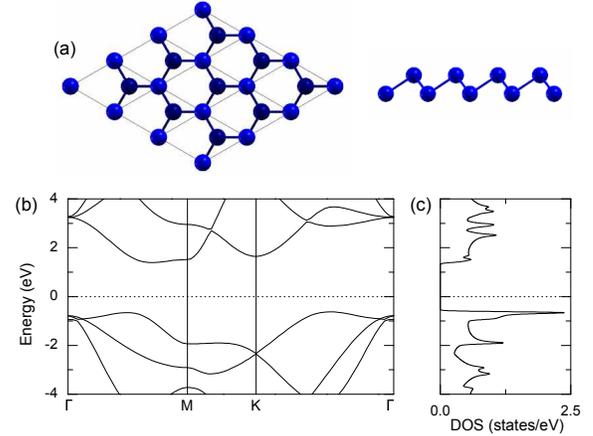}
\caption{(color online) (a) Top (left panel) and side (right panel) view of blue
  phosphorene. A 3$\times$3 supercell is taken for
  clarity. (b) Band structure and (c) density of states (DOS) for the
blue phosphorus monolayer. The Fermi energy level is set to 0 eV.}
\label{fig1}
\end{figure}

If we change the lattice structure of black phosphorus by little, the
situation is dramatic. This is the case of monolayer of blue
phosphorus (blue phosphorene), which is theoretically predicted through using {\it ab
  initio} method \cite{zhu}. 
It has an energy gap of 2
eV and thus has a great potential for practical applications. Thus, it
is helpful to study its electronic properties 
thoroughly to broaden their use. Experimentally, single-layer
phosphorus has been produced by exfoliating its bulk. If the monolayer
is further cut into nanoribbons, an increase of band gap would be
expected considering the quantum confinement effect. 
So far, the study of blue phosphorene nanoribbons (BPNRs) is still missing.

In this work, we show that BPNRs with hydrogen passivated armchair or
zigzag shaped edges have indirect band gaps. In addition, the gap
increases as the decrease of nanoribbon's width, which originates from
quantum confinement effect. By fitting the calculated gaps with
respect to their widths, we extract the quantum
confinement parameters. The results demonstrate that the quantum confinement
effect in armchair BPNRs is more pronounced than that in zigzag BPNRs. For
the case of armchair BPNRs, the quantum confinement mainly dominates
the lowest unoccupied states. In contrast, the quantum confinement
only affects the highest occupied states for zigzag BPNRs.

The rest of this work is arranged as follows. In Sec. II we briefly
describe the method used in this work. Results and discussion are
represented in Sec. III. Finally, we conclude our work in Sec. IV.

\section{II. Method}

Our first-principles calculations are performed through using the
SIESTA code within the framework of density functional theory (DFT)
\cite{siesta}. The generalized gradient approximation
Perdew-Burke-Ernzerhof (PBE) exchange-correction functional \cite{pbe}
and the norm-conserving pseudopotentials are used. The plane wave
energy cutoff is set to 210 Ry  to ensure the convergence of total
energy. The reciprocal space is sampled by a fine grid of
10$\times$1$\times$1 {\it k}-point in the Brillouin zone. The
conjugate gradient algorithm is taken to fully relax the geometry
until the force on each individual atom is less than 0.02
eV/{\AA}. The optimized double-$\zeta$ orbitals including
polarization orbitals are employed to describe the valence
electrons. Main results are checked using VASP code and a good
agreement between them is obtained.

\section{III. Results and discussion}

\begin{figure}
\includegraphics[width=7cm]{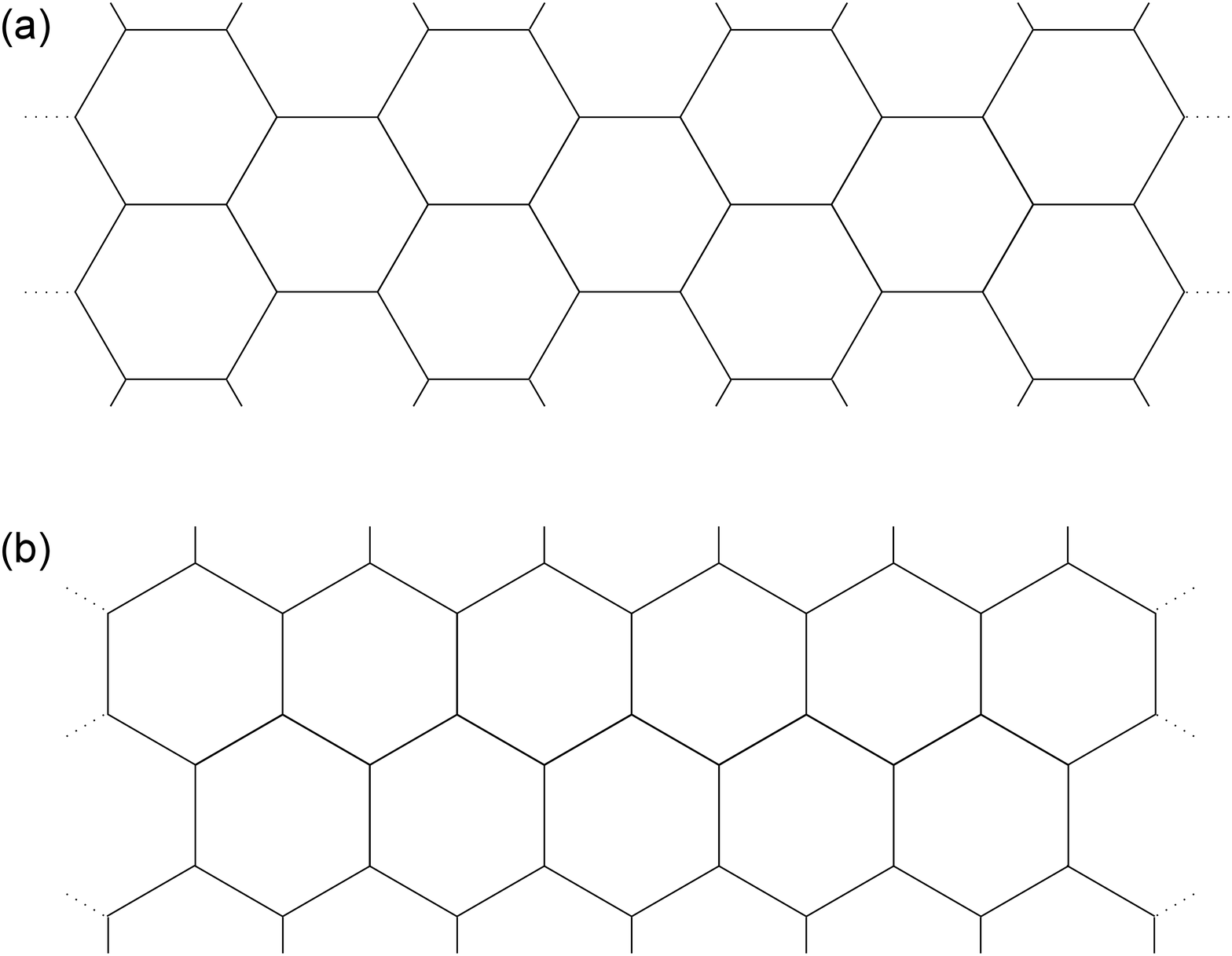}
\caption{(a) Armchair and (b) zigzag nanoribbons of blue phosphorene.
  For stabilizing the edge's state, the hydrogen saturation
  is taken in the realistic calculation. }
\label{fig2}
\end{figure}

As a start, it is necessary to test the equilibrium configuration of
blue phosphorene, as shown in Fig. \ref{fig1}(a). The obtained lattice
constant is 3.32 {\AA}. The P-P bond length is 2.29 {\AA}, which is
slightly larger than the P-P covalent bond length ($\sim$2.2 {\AA}) of
black phosphorus \cite{du}. The two
inequivalent P atoms are distributed in two planes with a separation of
1.26 {\AA} (see right panel of Fig. \ref{fig1}(a)). All the above 
lattice parameters agree well with the recent theoretical results
\cite{zhu}.  According to the band structure as shown in
Fig. \ref{fig1}(b), the blue phosphorene is indeed a
semiconductor. The highest occupied states appear at the middle region
along the K-$\Gamma$ line, while the lowest unoccupied states are
located between the $\Gamma$ and M points. It reveals 
an energy gap of $\sim$2 eV. It should be
noticed that our band structure has a slight difference from that
reported by Zhu {\it et al.} \cite{zhu}. In our case two occupied levels are nearly
degenerate at the K point below the Fermi energy level. By contrast, a
gap of around 0.5 eV appears in Zhu's work \cite{zhu}. At present, we 
can not understand such a divergence. The DOS of blue
phosphorene is displayed in Fig. \ref{fig1}(c), which matches the band
structure well. The very localized states appear at 1 eV below the
Fermi level, corresponding the highest occupied states.

\begin{figure}
\includegraphics[width=8cm]{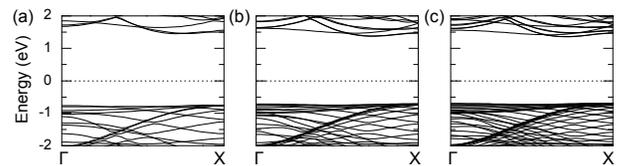}
\caption{Band structures of aBPNRs for three widths: (a) 11-aBPNR,
  (b) 19-aBPNR, and (c) 27-aBPNR. The Fermi energy level is set to 0
  eV.} 
\label{fig3}
\end{figure}

Upon cutting along different directions, two types of nanoribbons of
armchair and zigzag are generated, as illustrated in
Fig. \ref{fig2}. The same denotation as graphene nanoribbons \cite{son}
is taken for these BPNRs. Following
previous convention, the armchair BPNRs are classified by the number of dimer lines ($N_{a}$) across the
ribbon width. Likewise, zigzag ribbons
are sorted by the number of the zigzag chains ($N_{z}$) across the
ribbon width. We refer to a BPNR with $N_{a}$ dimer lines as a
$N_{a}$-aBNNR and a BPNR with $N_{z}$ zigzag chains as a
$N_{z}$-zBPNR.

\begin{figure}
\includegraphics[width=6cm]{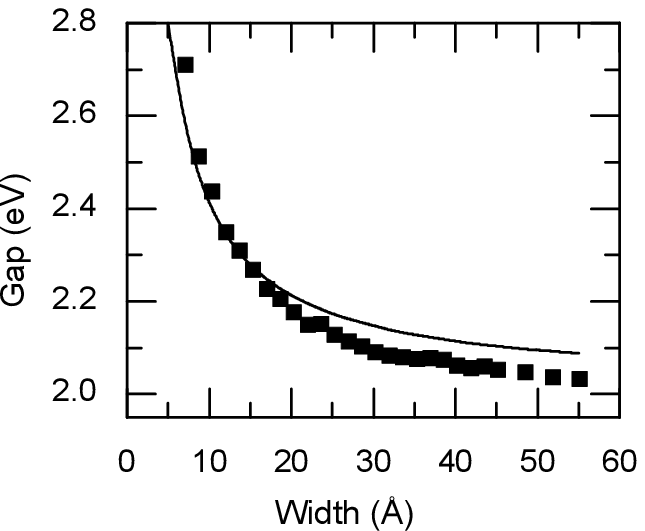}
\caption{The variation of energy gap as a function of width for
  aBPRBs. The squares are the calculated values. The
  solid line is the fitted curve.}
\label{fig4}
\end{figure}

It is well known that quantum confinement plays a key role 
when an infinite monolayer is cut into a nanoribbon. This quantum
confinement has a direct effect on energy gap \cite{niujap}. This is
also true in our case. 
Figure \ref{fig3} shows the calculated band structures of the aBPMRs for
three widths. Our finding is very insightful. As the width increases,
the highest occupied states remain unchanged. In contrast, the lowest
unoccupied states move down, which is a direct manifestation of quantum
confinement effect to aBPNRs.  In order to understand this quantum confinement
better, we plot the energy gaps as a function of widths, as shown in
Fig. \ref{fig4}. As can be seen that the energy gap is close to 2.7 eV
when the nanoribbon's width is about 10 {\AA}. This is a big
improvement compared to its infinite monolayer. As the width further
increases, the energy gap decreases to be $\sim$2 eV.

Next, we fit the theoretical gaps by $E = E_{0} + \gamma w^{-1}$ with $E_{0}$
being the energy gap of blue phosphorene, $w$ the width of
nanoribons in unit of {\AA}, and $\gamma$ the quantum confinement
parameter in unit of eV{\AA}. $\gamma$ represents the strength of quantum confinement and
$\gamma=0$ corresponds to non quantum confinement.
Here we obtain $\gamma = 3.93\pm 0.18$
eV{\AA} for aBPNRs. The corresponding fitted curve (solid line) is
given in Fig. \ref{fig4}, indicating a good agreement between the
calculated gaps and the fitted curve.

\begin{figure}
\includegraphics[width=6cm]{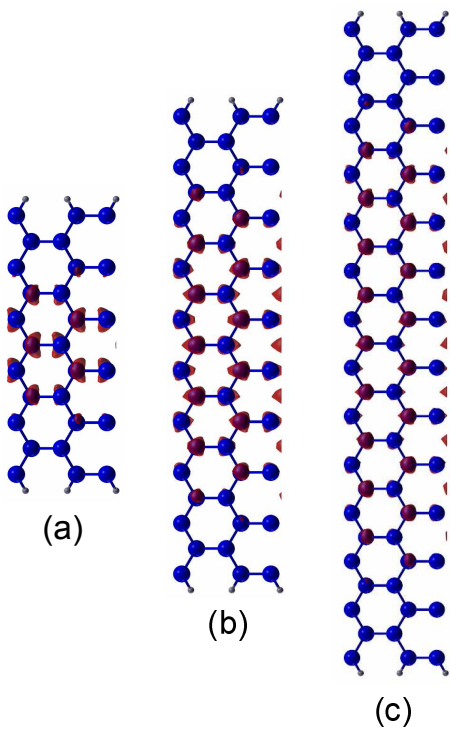}
\caption{(color online) Local density of states (LDOS) of the lowest unoccupied
  states under the same energy range above the Fermi energy level for
  three aBPNRs: (a) 11-aBPNR, (b) 19-aBPNR, and (c) 27-aBPNR. The
  isovalue is set to 0.0035 $|e|$/Bohr$^{3}$. The blue and white balls
  represent the P and H atoms, respectively.}
\label{fig5}
\end{figure}

To examine the underlying physics of this quantum confinement effect related
to the width of aBPNRs, we plot the local density of states (LDOS) of
the lowest unoccupied states under the same energy range above the
Fermi level, as shown in Fig. \ref{fig5}. 
 As the width is small, these LDOS locate
at the middle region of aBPNR (see Fig. \ref{fig5}(a)). They form
crescent moons, showing a strong covalent manner. As the width
increases, the LDOS shrink. In other words, the edge states without
LDOS become enhanced.  Interestingly, the LDOS seem to be spherical
(see Fig. \ref{fig5}(c)).  This means the covalent character become
weaker.

\begin{figure}
\includegraphics[width=8cm]{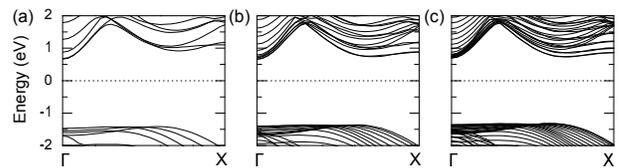}
\caption{Band structures of zBPNRs for three widths: (a) 6-zBPNR, (b)
  14-zBPNR, and (c) 22-zBPNR. The Fermi energy level is set to 0 eV.} 
\label{fig6}
\end{figure}

\begin{figure}
\includegraphics[width=6cm]{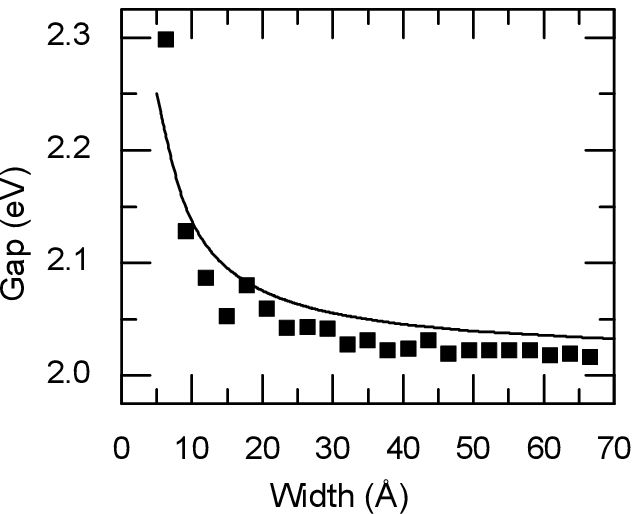}
\caption{The variation of energy gap as a function of width for
  zBPRBs. The squares are the calculated values. The
  solid line is the fitted curve.}
\label{fig7}
\end{figure}

In following, we switch our focus on zBPNRs. The band structures of
zBPNRs for three widths are shown in Fig. \ref{fig6}. They are also
indirect semiconductors where the highest occupied states are near the
$\Gamma$ point and the lowest unoccupied states are close to the X
point. It is also found that the widths have an effect on the band
structure. In contrast,
this quantum confinement only acts on the highest occupied states near the
$\Gamma$ point. As the width increases, the highest occupied states
are lifted up slightly, resulting in a decrease in energy gap.
The detailed variation of gaps as a function of widths
is displayed in Fig. \ref{fig7}. It explicitly shows that the energy
gap increases up to be $\sim$2.3 eV when the width of zBPNR
is about 8 {\AA}. If we use the above formula to fit these data, a
good agreement is obtained (see the solid line in
Fig. \ref{fig7}). The fitted quantum confinement parameter $\gamma$
equals to 1.38$\pm$0.16 eV{\AA}, which is much smaller than that value
(3.93$\pm$0.18 eV{\AA}) for aBPNRs. This implies that the quantum
confinement in aBPNRs is stronger than that in zBPNRs.

\begin{figure}
\includegraphics[width=5cm]{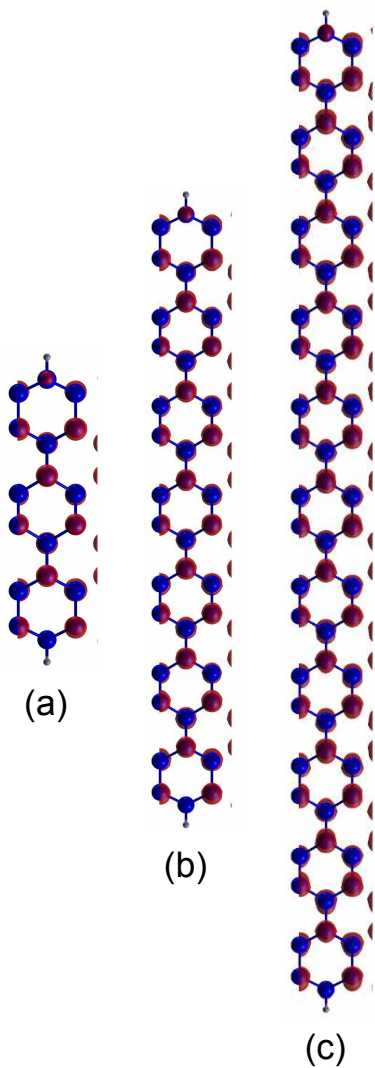}
\caption{(color online) The LDOS of the highest occupied states under the
  same energy range below 
the Fermi energy level for three zBPNRs: (a) 6-zBPNR, (b)
14-zBPNR, and (c) 22-zBPNR. The isovalue is set to 0.01
$|e|$/Bohr$^{3}$. The blue and white balls represent the P and H
atoms, respectively. }
\label{fig8}
\end{figure}

The corresponding LDOS are displayed in Fig. \ref{fig8}. In contrast to
aBPNRs, they distribute evenly on each P atoms.  As the width is
small, the LDOS are spherical (see Fig. \ref{fig8}(a)), showing a very
localized feature. This makes the occupied states far away the Fermi
energy level (see Fig. \ref{fig6}). 
As the width further increases, the LDOS exhibit a slight
polarization (see Fig. \ref{fig8}(c) for more details), showing a more
dispersion of electronic states. As a result, the highest occupied
states will be lifted up. This is the origin of the band gap evolution
with the widths of zBPNRs.

\section{IV. conclusion}

In conclusion, we have shown that blue phosphorene nanoribbons with
 armchair or zigzag shaped edges all have energy gaps, which
decrease as the width of the system increase. The role of the quantum
confinement or the width is crucial for the values for the band gaps.

\section*{Acknowledgments}

This work was supported by the National Basic Research Program of
China under Grant No. 2012CB933101 and  the National Science
Foundation under Grant No. 51372107. M.S.S. thanks the State
Scholarship Fund by the China Scholarship Council for financially
supporting his visit to Indiana State University.

$^{*}$Email: sims@lzu.edu.cn

\end{document}